\documentclass[aip, twocolumn, reprint,jcp]{revtex4-2} %superscriptaddress,eqsecnum,

\usepackage{amsmath,amssymb,amsfonts,bm}

\usepackage{graphicx,epstopdf}  %epsfig,
\usepackage{color}
\usepackage{extarrows}
\usepackage{enumitem}
\usepackage{empheq}
\usepackage{tensor}

\newcommand{\ud}{{\mathrm d}}

\newcommand{\Sch}{Schr\"{o}dinger\ }

\newcommand{\w}{\omega}

\newcommand{\B}{{\mbox{\tiny B}}}

\newcommand{\T}{{\mbox{\tiny T}}}
\newcommand{\tK}{{\mbox{\tiny K}}}
\newcommand{\tS}{{\mbox{\tiny S}}}

\newcommand{\SB}{\mbox{\tiny SB}}

\newcommand{\dg}{\dagger}
\newcommand{\la}{\langle}
\newcommand{\ra}{\rangle}

\newcommand{\Sec}[1]{Sec.\,\ref{#1}}

\newcommand{\nl}{\nonumber \\}
\newcommand{\be}{\begin{equation}}
\newcommand{\ee}{\end{equation}}
\newcommand{\bsube}{\begin{subequations}}
\newcommand{\esube}{\end{subequations}}
\newcommand{\Eq}[1]{Eq.\,(\ref{#1})}
\newcommand{\Eqs}[1]{Eqs.\,(\ref{#1})}
\newcommand{\Fig}[1]{Fig.\,\ref{#1}}

\newcommand{\RN}[1]{%
  \textup{\uppercase\expandafter{\romannumeral#1}}%
}

\allowdisplaybreaks[1]

\usepackage{hyperref}  
\hypersetup{hidelinks,
	colorlinks=true,
	allcolors=black,
	pdfstartview=Fit,
	breaklinks=true
}

\begin{document}

\title{Extended dissipaton equation of motion for  electronic open quantum systems:\\ Application to the Kondo impurity model 
}
%%%
\author{Yu Su}
\affiliation{Hefei National Research Center for Physical Sciences at the Microscale,  University of Science and Technology of China, Hefei, Anhui 230026, China}
\author{Zi-Hao Chen}
\affiliation{CAS Key Laboratory of Precision and Intelligent Chemistry, University of Science and Technology of China, Hefei, Anhui 230026, China
}
\author{Yao Wang}
\email{wy2010@ustc.edu.cn}
\affiliation{Hefei National Research Center for Physical Sciences at the Microscale,  University of Science and Technology of China, Hefei, Anhui 230026, China}
\author{Xiao Zheng}
\affiliation{Hefei National Research Center for Physical Sciences at the Microscale,  University of Science and Technology of China, Hefei, Anhui 230026, China}
\affiliation{Department of Chemistry, Fudan University, Shanghai 200433, China}
\author{Rui-Xue Xu}
\affiliation{Hefei National Research Center for Physical Sciences at the Microscale,  University of Science and Technology of China, Hefei, Anhui 230026, China}
\affiliation{CAS Key Laboratory of Precision and Intelligent Chemistry, University of Science and Technology of China, Hefei, Anhui 230026, China
}
\author{YiJing Yan}
\email{yanyj@ustc.edu.cn}
\affiliation{Hefei National Research Center for Physical Sciences at the Microscale,  University of Science and Technology of China, Hefei, Anhui 230026, China}
\affiliation{Department of Chemistry, School of Science, Westlake University, Hangzhou, Zhejiang 310024, China}

\date{\today}

\begin{abstract}
In this paper, we present an extended dissipaton equation of motion for studying the dynamics of electronic impurity systems.
Compared with the original theoretical formalism, 
the quadratic couplings are  
introduced into the  Hamiltonian accounting for the interaction between the impurity and its surrounding environment.
By exploiting the quadratic fermionic dissipaton algebra, the proposed extended dissipaton equation of motion offers a powerful tool for studying the dynamical behaviors of electronic impurity systems, particularly in situations where nonequilibrium and strongly correlated effects play significant roles.
Numerical demonstrations are carried out to investigate the temperature dependence of the Kondo resonance in the Kondo impurity model. 
\end{abstract}
\maketitle

\section{Introduction}
Electronic impurity systems are important in a wide range of fields, including solid--state physics, materials science, quantum information, and so on. 
\cite{Kon68437,Sma83515,Zit06045312,Bal06373,Foo15103112}
The dynamics of these systems are particularly intriguing due to the strong coupling between the impurity and its surrounding environment.
\cite{Her913720,Din9715521,Din97R15521,Sch9814978,Dic014505,Luo04256602,Hur07126801,Li982893,Rui14035119}
The study of electronic impurity systems is crucial for understanding the behavior of materials and quantum devices, and has practical implications for designing new technologies. 
\cite{Ujs002557,Ham04115313,Sch09235130,Isi10235120,Coh11075150,Tsu11187201,Ort12125324,Zha17075410}

One of the main challenges in studying electronic impurity systems is accurately modeling their interactions with the environment. 
{\color{black}
The Anderson and Kondo impurity models are widely used for describing the
impurity system within the fermionic environments.}
The Anderson model describes a local quantum impurity coupled to non-interacting conduction electrons in a metal, where the impurity system is represented by a single electronic level interacting with a continuum of reservoir states.
The system--bath coupling is in the \emph{linear} form with respect to the creation and annihilation operators of the impurity and bath states,
$H_{\SB}\sim \sum_{ks}(t_{ks}\hat d_{ks}^{\dg}\hat a_{s}+{\rm h.c.})$.
{\color{black}
The other model, the Kondo
impurity model, is famous for successfully predicting the emergence of a many--body state at low temperatures, known as the Kondo resonance, which is
featured as a sharp peak in the vicinity of the Fermi level of the metal electrons.
The Kondo impurity model is analogous to the Anderson impurity model,
describing an impurity spin coupled to conduction electrons in a metal. But the
electron--electron interactions take a Heisenberg coupling form, $H_{\SB}\sim J {\bm S}_{\rm imp} \cdot {\bm S}_{\B}$,
where $J$ is the exchange coupling constant between the impurity spin ${\bm S}_{\rm imp}$ and the conduction electrons total spin ${\bm S}_{\B}$, and  ${\bm S}_{\B}$ is \emph{quadratic} with respect to the reservoir creation and annihilation operators.
}

So far, various methods have targeted the equilibrium and dynamical properties of quantum impurities, such as the  quantum Monte Carlo method, \cite{Gul11349,Har15085430} the numerical renormalization group method \cite{Wil75773,Yos909403,Cos973003,Bul988365,Pet06245114,And08195216} and its time--dependent extension, \cite{And05196801,Fri06144410} the time--dependent density matrix renormalization group method, \cite{Nis04613,Mer12075153}, and so on.
{\color{black}
The recent developments in methods include
the time evolving density matrices using orthogonal polynomials algorithm (TEDOPA),
\cite{Pri10050404, Nus20155134}
the time-evolving matrix product operator (TEMPO) algorithm, 
\cite{Jor19240602,Ric22167403}
 the automated compression of environments (ACE) method, \cite{Cyg22662} the inchworm quantum Monte Carlo method,
\cite{Coh15266802, Che17054105, Cai202430} 
  the quantum quasi-Monte Carlo algorithm, \cite{Ber21155104} and the auxiliary master equation approach (AMEA). \cite{Sor19043303}
}

Especially, as a time--derivative equivalence to the Feynman--Vernon influence functional path,\cite{Fey63118} the hierarchical equations of motion (HEOM) method has attracted increasing attention, with either bosonic \cite{Tan20020901,Tan89101,Tan906676,Yan04216,Tan06082001,Xu05041103,Xu07031107,Din12224103} or fermionic bath environment influence. \cite{Jin08234703,Li12266403,Ye16608} 
{\color{black}
Earlier applications of the HEOM method have been mainly
focused on the Anderson impurity model, because the method is developed on
the basis of a linear system--bath coupling scenario. 
The extension to considering also the quadratic system--bath coupling form, which is the case for the Kondo impurity model, is yet to be developed. 
The Kondo impurity model has so far been dealt with by such as the renormalization group approach.\cite{Wil75773,Cos001504,Fri06144410}
}

Dissipaton equation of motion (DEOM),\cite{Yan14054105, Wan22170901} as a second quantization version of HEOM, is able to acquire the dynamics in the presence of nonlinear coupling in the bosonic scenarios. \cite{Xu17395,Xu18114103}
Its exactness has been numerically verified recently.\cite{Che23074102}
{\color{black} In this work, we propose  the fermionic version of the extended DEOM (\textrm{ext}-DEOM) for the fermionic quadratic coupling between the system and bath.}
This addresses the challenge of DEOM to deal with the Kondo impurity model, where the quadratic couplings between the impurity and its environment are involved.
This extension builds upon previously developed fermionic dissipaton algebra introduced for linear couplings and expands the capabilities to quadratic environment coupling scenarios.

The remainder of this paper is organized as follows: In \Sec{thsec:deom}, we propose the ext-DEOM with a detailed derivation. In \Sec{sec:num}, we  demonstrate the temperature-dependent Kondo resonance in the Kondo impurity model. Finally, we summarize our paper in \Sec{sum}. Throughout this paper, we set $\hbar  =1$ and $\beta = 1/(k_BT)$, with $k_B$ being the Boltzmann constant and $T$ the temperature.

\section{Extended dissipaton equation of motion}\label{thsec:deom}

\subsection{Quadratic system--bath interactions}

{\color{black} In this work, we consider an electronic system ($H_{\tS}$) in contact with a
fermionic bath ($h_{\B}$).}
While $H_{\tS}$ is arbitrary, the bath Hamiltonian $h_{\B}$ is modeled as noninteracting electrons,
\begin{align}\label{hB}
  h_{\B} = \sum_{ks} \epsilon_{ks}\hat d^+_{ks}\hat d_{ks}^-,
\end{align}
where $k$ and $s=\uparrow,\downarrow$ label a single--electron spin--orbital state. The system and bath couple with each other via the quadratic interaction, 
\begin{align}\label{SB2}
  H_{\SB} = \frac{1}{2}\sum_{\sigma us}\sum_{\sigma' vs'}\hat q^{\bar\sigma\bar\sigma'}_{us,vs'}\hat{\Phi}^{\sigma}_{us}\hat{\Phi}^{\sigma'}_{vs'}.
\end{align}
Here, $\sigma\in\{+,-\}$ and the hybridizing bath operators read 
\begin{align}\label{Phi}
  \hat\Phi^+_{us} \equiv \sum_k c_{kus}
{\color{black}  
  \hat d^+_{ks}
}  
   \equiv (\hat\Phi^-_{us})^\dagger.
\end{align}
$\{\hat q^{\sigma\sigma'}_{us,vs'}\}$ are the system subspace operators, generally quadratic in terms of the system creation/annihilation operators $\{\hat a_{us}^\sigma\}$. It is closely related to the form of two--particle interactions in many--electron systems. Without loss of generality, the $\{\hat q^{\sigma\sigma'}_{us,vs'}\}$ assume antisymmetric, 
\begin{align}\label{qanti}
  \hat q^{\sigma\sigma'}_{us,vs'}=-\hat q^{\sigma'\sigma}_{vs',us}.
\end{align}
% These settings will be specified later in \Sec{2B} based on concrete physical models.

\subsection{Fermionic bath statistics and dissipaton decomposition}

For the environment given by \Eqs{hB} and (\ref{Phi}), the hybridizing bath spectral density functions can completely describe the bath influence, defined as \cite{Yan14054105,Yan16110306} 
{\color{black}
\begin{align}
  \Gamma_{uvs}(\omega) \equiv \Gamma^-_{uvs}(\omega) = \pi\sum_{k}c_{kus}^*c_{kvs}\delta(\omega-\epsilon_{ks}).
\end{align}
}
It can be equivalently expressed via 
{\color{black}
\begin{align}\label{eq6}
  \Gamma^\sigma_{uvs}(\omega) \equiv \frac{1}{2}\int_{-\infty}^\infty\!\!\ud t\, e^{-\sigma i\omega t}\la\{ \hat \Phi^\sigma_{us}(t),\hat\Phi^{\bar\sigma}_{vs}(0) \}\ra_{\B},
\end{align}
}
with {\color{black}$\Gamma^\sigma_{vus}(\omega)=[\Gamma^\sigma_{uvs}(\omega)]^*=\Gamma^{\bar\sigma}_{uvs}(\omega)$}. Here, we follow the bare--bath thermodynamic prescription: $\hat\Phi^\sigma_{us}(t)\equiv e^{ih_{\B}t}\hat\Phi^\sigma_{us}e^{-ih_{\B}t}$ and $\la\hat O\ra_{\B}\equiv {\rm tr}_{\B}(\hat Oe^{-\beta h_{\B}})/{\rm tr_{\B}}(e^{-\beta h_{\B}})$. We then have 
\begin{align}\label{FDT}
  \la\hat\Phi^\sigma_{us}(t)\hat\Phi^{\bar\sigma}_{vs}(0)\ra_{\B} = \frac{1}{\pi}\int_{-\infty}^\infty\!\!\ud\omega\, e^{\sigma i\omega t}\frac{{\color{black}\Gamma^\sigma_{uvs}(\omega)}}{1+e^{\sigma\beta\omega}}.
\end{align}
This is the fermionic fluctuation--dissipation theorem.\cite{Yan16110306} 

{\color{black}
Generally, the
influence of the bath on the system dynamics in this case [cf.\ \Eq{SB2}] should be encoded in the
fourth and higher order correlation functions, such as $\la\hat\Phi^{\sigma_1}_{u_1s_1}(t_1)\hat\Phi^{\sigma_2}_{u_2s_2}(t_2)\hat\Phi^{\sigma_3}_{u_3s_3}(t_3)\hat\Phi^{\sigma_4}_{u_4s_4}(t_4)\ra_{\B}$.
However, since the bare--bath thermodynamic prescription [cf.\,the description below \Eq{eq6}] and the noninteracting electrons model [cf.\,\Eq{hB}], all fourth and higher order correlations can be decomposed into the product of second order ones in \Eq{FDT}.  
This is known as the Bloch--de Dominicis theorem. \cite{Bog09}
}

To proceed, we expand \cite{Yan14054105}
\begin{align}\label{Fc}
  \la\hat\Phi^\sigma_{us}(t)\hat\Phi^{\bar\sigma}_{vs}(0)\ra_{\B} = \sum_{\kappa=1}^K{g}^\sigma_{\kappa uvs}e^{-\gamma^\sigma_{\kappa uvs}t}.
\end{align}
Its time reversal reads
\begin{align}\label{Fcc}
  \la\hat\Phi^{\bar\sigma}_{vs}(0)\hat\Phi^\sigma_{us}(t)\ra_{\B} = \sum_{\kappa=1}^K{g}^{\bar\sigma*}_{\kappa uvs}e^{-\gamma^\sigma_{\kappa uvs}t},
\end{align}
with $\gamma^\sigma_{\kappa uvs}=(\gamma^{\bar\sigma}_{\kappa uvs})^*$ required. We can then decompose 
\begin{align}
  \hat{\Phi}^\sigma_{us} = \sum_{\kappa=1}^K\hat\phi^\sigma_{\kappa us},
\end{align}
with 
\begin{subequations}\label{fcf}
  \begin{align}\label{fcf1}
    &\la\hat\phi^\sigma_{\kappa us}(t)\hat\phi^{\sigma'}_{\kappa' vs'}(0)\ra_{\B}\! = \!\delta^{\sigma\bar\sigma'}_{\kappa s,\kappa's'}{g}^\sigma_{\kappa uvs}e^{-\gamma^\sigma_{\kappa uvs}t},\\
    \label{fcf2}
    &\la\hat\phi^{\sigma'}_{\kappa'vs'}(0)\hat\phi^\sigma_{\kappa us}(t)\ra_{\B}\! =\! \delta^{\sigma\bar\sigma'}_{\kappa s,\kappa's'}{g}^{\bar\sigma*}_{\kappa uvs}e^{-\gamma^{\sigma}_{\kappa uvs}t}.
  \end{align}
\end{subequations}
Here, $\{\hat\phi^\sigma_{\kappa us}\}$ are denoted as the dissipaton operators, providing a statistical quasi--particle picture to account for the Gaussian environmental influences. It is evident that \Eq{fcf} can reproduce both \Eqs{Fc} and (\ref{Fcc}). 

For simplicity, we adopt the index abbreviations,
\begin{align}
  j\equiv (\sigma\kappa us)\ \ \text{and}\ \ \bar j\equiv(\bar\sigma\kappa us),
\end{align}
leading to $\hat\phi_j\equiv \hat \phi^\sigma_{\kappa us}$ and so on. Then we can recast \Eq{SB2} as
\begin{align}\label{SB2f}
  H_{\SB} = \frac{1}{2}\sum_{jj'}\hat q_{\bar j\bar j'}\hat \phi_j\hat \phi_{j'}.
\end{align}
Here, we define $\hat q_{j j'}\equiv \hat q^{\sigma\sigma'}_{us,vs'}$.

\subsection{Extended fermionic DEOM formalism}

Dissipaton operators, together with the total system density operator $\rho_{\T}(t)$, form the dynamical variables of DEOM, namely the dissipaton density operators (DDOs),\cite{Yan14054105}
\begin{align}
  \rho_{\bf j}^{(n)}(t)\equiv \rho_{j_1\cdots j_n}^{(n)}(t)\equiv {\rm tr}_{\B}\big[(\hat \phi_{j_n}\cdots \hat \phi_{j_1})^\circ\rho_{\T}(t)\big].
\end{align}
The notation, $(\cdots)^\circ$, denotes the \textit{irreducible} dissipaton product notation, with $(\hat \phi_j\hat \phi_{j'})^\circ=-(\hat \phi_{j'}\hat \phi_j)^\circ$ for fermionic dissipatons.
{\color{black}
Note that the reduced system density operator is $\rho^{(0)}_{}(t)={\rm tr}_{\B}[\rho_{\T}(t)]\equiv \rho_{\tS}(t)$. 
}

{\color{black} In the dissipaton
theory, we assume} ($i$) Each dissipaton satisfies the generalized diffusion equation,\cite{Yan14054105,Yan16110306}
\begin{align}\label{gdiff}
  {\rm tr}_{\B}\bigg[ \bigg( \frac{\partial}{\partial t}\hat \phi_j \bigg)_{\B}\rho_{\T}(t) \bigg] = -\gamma_j{\rm tr}_{\B}[\hat \phi_j\rho_{\T}(t)],
\end{align}
where $(\frac{\partial}{\partial t}\hat \phi_j)_{\B} = -i[\hat \phi_j,h_{\B}]$. \Eq{gdiff} arises from that each dissipaton is associated with a single exponent, for its forward and backward correlation functions [cf.\,(\ref{fcf})]. ($ii$) The generalized Wick's theorems (GWT) deal with adding dissipaton operators into the irreducible notation. The GWT-1s evaluate the linear bath coupling with one dissipaton added each time. They are expressed as \cite{Yan14054105,Yan16110306} 
\begin{subequations}\label{GWT1}
  \begin{align}
    &\quad\, {\rm tr}_{\B}\big[ (\hat \phi_{j_n}\cdots\hat \phi_{j_1})^\circ\hat \phi_j\rho_{\T}(t) \big]\nl 
    & = \rho_{j{\bf j}}^{(n+1)}(t)+\sum_{r=1}^n(-)^{r-1}\la\hat \phi_{j_r}\hat \phi_j\ra^>_{\B}\rho_{{\bf j}_r^-}^{(n-1)}(t)
  \end{align}
  and
  \begin{align}
    &\quad\, {\rm tr}_{\B}\big[ \hat \phi_j(\hat \phi_{j_n}\cdots\hat \phi_{j_1})^\circ\rho_{\T}(t) \big]\nl 
    & = \rho_{{\bf j}j}^{(n+1)}(t)+\sum_{r=1}^n(-)^{n-r}\la\hat \phi_j\hat \phi_{j_r}\ra^<_{\B}\rho_{{\bf j}_r^-}^{(n-1)}(t),
  \end{align}
\end{subequations}
where we denote $\la\hat \phi_{j}\hat \phi_{j'}\ra^>_{\B} \equiv \la\hat \phi_{j}(0+)\hat \phi_{j'}\ra_{\B}$, $\la\hat \phi_{j'}\hat \phi_{j}\ra^<_{\B} \equiv \la\hat \phi_{j'}\hat \phi_{j}(0+)\ra_{\B}$, and ${\bf j}_r^- \equiv \{ j_n\cdots j_{r+1}j_{r-1}\cdots j_1 \}$. Moreover, the GWT-2s are similarly given by 
\begin{subequations}\label{GWT2}
  \begin{align}\label{GWT-1}
    &\quad\,{\rm tr}_{\B}\big[(\hat \phi_{j_n}\cdots\hat \phi_{j_1})^\circ\hat \phi_j\hat \phi_{j'}\rho_{\T}(t)\big]\nl 
  &= \rho_{j'j{\bf j}}^{(n+2)}(t)+\la\hat \phi_j\hat \phi_{j'}\ra_{\B}\rho_{\bf j}^{(n)}(t)\nl 
  &\quad -\sum_{r=1}^n(-)^{r-1}\la\hat \phi_{j_r}\hat \phi_{j'}\ra_{\B}^>\rho_{j{\bf j}_{r}^-}^{(n)}(t)\nl 
  &\quad +\sum_{r=1}^n(-)^{r-1}\la\hat \phi_{j_r}\hat \phi_j\ra^>_{\B}\rho_{j'{\bf j}_r^-}^{(n)}(t)\nl 
  &\quad +\sum_{r,r'}\theta_{rr'}\la\hat \phi_{j_r}\hat\phi_j\ra_{\B}^>\la\hat \phi_{j_{r'}}\hat \phi_{j'}\ra_{\B}^>\rho_{{\bf j}_{rr'}^{--}}^{(n-2)}(t)
  \end{align}
  and 
  \begin{align}\label{GWT-2}
  &\quad\, {\rm tr}_{\B}\big[\hat \phi_j\hat \phi_{j'}(\hat \phi_{j_n}\cdots\hat \phi_{j_1})^\circ\hat \rho_{\T}(t)\big]\nl 
  &= \rho_{{\bf j}j'j}^{(n+2)}(t)+\la\hat \phi_j\hat \phi_{j'}\ra_{\B}\rho_{\bf j}^{(n)}(t)\nl 
  &\quad -\sum_{r=1}^n(-)^{n-r}\la\hat \phi_{j}\hat \phi_{j_r}\ra_{\B}^<\rho_{{\bf j}_{r}^-{j'}}^{(n)}(t)\nl 
  &\quad +\sum_{r=1}^n(-)^{n-r}\la\hat \phi_{j'}\hat \phi_{j_r}\ra^<_{\B}\rho_{{\bf j}_r^-j}^{(n)}(t)\nl 
  &\quad -\sum_{r,r'}\theta_{rr'}\la\hat \phi_{j'}\hat \phi_{j_r}\ra_{\B}^<\la\hat \phi_j\hat \phi_{j_{r'}}\ra_{\B}^<\rho_{{\bf j}_{rr'}^{--}}^{(n-2)}(t).
  \end{align}
\end{subequations}
Here, 
\begin{align}\label{eq18}
  \theta_{rr'} \equiv 
  \begin{cases}
    (-)^{r-r'}, \ \ & r \geq r' \\
    (-)^{r-r'+1}, \ \ & r < r' \\
  \end{cases}
\end{align}
and ${\bf j}_{rr'}^{--} \equiv \{ j_n\cdots j_{r+1}j_{r-1}\cdots j_{r'+1}j_{r'-1} \cdots j_1 \} = {\bf j}_{r'r}^{--}$.

Then, by applying the dissipaton algebras on the von Neumann--Liouville Equation,
\begin{align}\label{liou}
  \dot{\rho}_{\T} = -i[H_{\T},\rho_{\T}] = -i[H_{\tS}+h_{\B}+H_{\SB},\rho_{\T}],
\end{align}
one can construct the \textrm{ext}-DEOM. We then, term by term, evaluate the contributions of specific {\color{black}three} components in the $H_{\T}$.
\begin{enumerate}[fullwidth,label=(\emph{\rm\alph*})]
  \item The $H_{\tS}$--contribution: Evidently, 
  \begin{align}\label{eq20}
    {\rm tr}_{\B}\Big\{(\hat \phi_{j_n}\cdots\hat \phi_{j_1})^\circ[H_{\tS},\rho_\T]\Big\} = [H_{\tS},\rho_{\bf j}^{(n)}].
  \end{align}
  \item The $h_{\B}$--contribution: Using \Eq{gdiff}, we have 
  \begin{align}\label{eq21}
    i{\rm tr}_{\B}\Big\{ (\hat \phi_{j_n}\cdots\hat \phi_{j_1})^\circ [h_{\B},\rho_{\T}] \Big\} = \gamma^{(n)}_{\bf j}\rho_{\bf j}^{(n)},
  \end{align}
  with $\gamma^{(n)}_{\bf j}\equiv \sum_{r=1}^n\gamma_{j_r}$.
  % Here, we used \Eqs{fcf} and (\ref{GWT1}). Note the partial trace cyclic rule for fermionic operators,
  % \begin{align}
  %   {\rm tr}_{\B}[\hat O^{(n)}\rho_{\T}\hat f_j] = (-)^{n+1}{\rm tr_{\B}}[\hat f_j\hat O^{(n)}\rho_{\T}]
  % \end{align}
  % with $\hat O^{(n)}$ containing $n$ anticommute operators.
  \item The $H_{\SB}$--contribution: By applying \Eqs{GWT-1} and (\ref{GWT-2}), we can readily obtain
  \begin{align}\label{eq22}
    &\quad {\rm tr}_{\B}\Big\{ (\hat \phi_{j_n}\cdots\hat \phi_{j_1})^\circ [H_{\SB},\rho_{\T}] \Big\}\nl 
    &= \frac{1}{2}\sum_{jj'}\hat q_{\bar j\bar j'}{\rm tr}_{\B}\big[ (\hat \phi_{j_n}\cdots\hat \phi_{j_1})^\circ \hat \phi_j\hat \phi_{j'}\rho_{\T} \big]\nl &
    \quad -\frac{1}{2}\sum_{jj'}{\rm tr}_{\B}\big[ (\hat \phi_{j_n}\cdots\hat \phi_{j_1})^\circ \rho_{\T}\hat \phi_j\hat \phi_{j'} \big]\hat q_{\bar j\bar j'}\nl
    &= \frac{1}{2}\sum_{jj'}[\hat q_{\bar j\bar j'},\rho_{{\bf j}j'j}^{(n+2)}]+\frac{1}{2}\sum_{\sigma us}\sum_{\sigma' vs'}\la\hat\Phi^\sigma_{us}\hat\Phi^{\sigma'}_{vs'}\ra_{\B}[\hat q^{\bar\sigma\bar\sigma'}_{us,vs'},\rho_{\bf j}^{(n)}]\nl 
    &\quad +\sum_{rvj}(-)^{n-r}\Big[ {g}^{\sigma_r}_{\kappa_ru_rvs_r}\hat{q}^{\sigma_r\bar\sigma}_{vs_r,us}\rho_{{\bf j}_r^-j}^{(n)}+{g}^{\bar\sigma_r*}_{\kappa_ru_rvs_r}\rho_{{\bf j}_r^-j}^{(n)}\hat{q}^{\sigma_r\bar\sigma}_{vs_r,us} \Big]\nl 
    &\quad +\sum_{r>r'}\sum_{uv}(-)^{r-r'}\Big[ {g}^{\sigma_r}_{\kappa_ru_rus_r}{g}^{\sigma_{r'}}_{\kappa_{r'}u_{r'}vs_{r'}}\hat{q}^{\sigma_r\sigma_{r'}}_{us_r,vs_{r'}}\rho_{{\bf j}_{rr'}^{--}}^{(n-2)}\nl 
    &\qquad -{g}^{\bar\sigma_r*}_{\kappa_ru_rus_r}{g}^{\bar\sigma_{r'}*}_{\kappa_{r'}u_{r'}vs_{r'}}\rho_{{\bf j}_{rr'}^{--}}^{(n-2)}\hat{q}^{\sigma_r\sigma_{r'}}_{us_r,vs_{r'}}\Big].
  \end{align}
\end{enumerate}

{\color{black}
To derive \Eq{eq22}, we use the form of $H_{\SB}$ in \Eq{SB2f} in the first equality. For the second equality, we use \Eqs{GWT-1} and (\ref{GWT-2}) with \Eq{eq18}, by further noting: 

\noindent(a) With $j\equiv (\sigma\kappa us)$ and $j'\equiv (\sigma'\kappa' v s')$, we have [cf.\,\Eq{fcf}]
\begin{align*}
\la\hat \phi_{j}\hat \phi_{j'}\ra^>_{\B}=\delta^{\sigma\bar\sigma'}_{\kappa s,\kappa's'}{g}^\sigma_{\kappa uvs}\ \ \text{and}\ \
\la\hat \phi_{j'}\hat \phi_{j}\ra^<_{\B}=\delta^{\sigma\bar\sigma'}_{\kappa s,\kappa's'}{g}^{\bar\sigma*}_{\kappa uvs}.
\end{align*}
Apparently, $\sum_{\kappa,\kappa'}\la\hat\phi_j\hat\phi_{j'}\ra_{\B} = \la\hat\Phi^\sigma_{us}\hat\Phi^{\sigma'}_{vs'}\ra_{\B}$.

\noindent (b) To obtain the last two terms related to $\{\rho_{{\bf j}_r^-j}^{(n)}\}$ and $\{\rho_{{\bf j}_{rr'}^{--}}^{(n-2)}\}$ in the second equality, we have to use the the antisymmetric property of $\{\hat q^{\sigma\sigma'}_{us,vs'}\}$ [cf.\,\Eq{qanti}] and ${\bf j}_{rr'}^{--}={\bf j}_{r'r}^{--}$ [cf.\,the notation explanation below \Eq{eq18}].
%
% When dealing with the last three terms in both \Eqs{GWT-1} and (\ref{GWT-2}), we have used the antisymmetric property of $\{\hat q^{\sigma\sigma'}_{us,vs'}\}$ [cf.\,\Eq{qanti}] and ${\bf j}_{rr'}^{--}={\bf j}_{r'r}^{--}$ [cf.\,the notation explanation below \Eq{eq18}].

Therefore, \Eq{liou} together with \Eqs{eq20}--(\ref{eq22}) leads to the final \textrm{ext}-DEOM formalism, which reads 
}
\begin{align}\label{eDEOM}
  \dot{\rho}_{\bf j}^{(n)} &= -\Big(i\mathcal{L}_{\tS}^{\rm eff}+\gamma_{\bf j}^{(n)}\Big)\rho_{\bf j}^{(n)}-i\sum_{r=1}^n\sum_{j}(-)^{n-r}\mathcal{B}_{j_rj}\rho_{{\bf j}_r^-j}^{(n)}\nl 
  &\quad -\frac{i}{2}\sum_{jj'}\tensor{\mathcal{A}}{_{\bar j\bar j'}}\rho_{{\bf j}j'j}^{(n+2)}-i\sum_{r>r'}(-)^{r-r'}\tensor{\mathcal{C}}{_{j_rj_{r'}}}\rho_{{\bf j}_{rr'}^{--}}^{(n-2)},
\end{align}
where the superoperators in $\{\rho^{(n)}\}$ parts are defined as
\begin{align}
  \mathcal{L}_{\tS}^{\rm eff}\hat O &\equiv [H_{\tS}+\la H_{\SB}\ra_{\B},\hat O], \\ 
  \!\!\!\!\!\mathcal{B}^{\sigma,\sigma'}_{\kappa us,u's'}\hat O &\equiv \sum_v \Big({g}^{\sigma}_{\kappa uvs}\hat{q}^{\sigma\bar\sigma'}_{vs,u's'}\hat O+{g}^{\bar\sigma*}_{\kappa uvs}\hat O\hat{q}^{\sigma\bar\sigma'}_{vs,u's'}\Big),
\end{align}
and actions on the $\{\rho^{(n\pm 2)}\}$ parts are given by
\begin{subequations}
  \begin{align}
    \tensor*{\mathcal A}{^{\sigma,\sigma'}_{us,vs'}}\hat O &\equiv [\hat{q}^{\sigma\sigma'}_{us,vs'},\hat O],\\
    \tensor*{\mathcal C}{_{\kappa us,\kappa'vs'}^{\sigma,\sigma'}}\hat O &\equiv \sum_{u'v'}\Big({g}^{\sigma}_{\kappa uu' s}{g}^{\sigma'}_{\kappa'vv's'}\hat{q}^{\sigma\sigma'}_{u's,v's'}\hat O\nl 
    &\qquad \quad- {g}^{\bar\sigma*}_{\kappa uu's}{g}^{\bar\sigma'*}_{\kappa'vv's'}\hat O\hat{q}^{\sigma\sigma'}_{u's,v's'}\Big).
  \end{align}
\end{subequations}

\section{Numerical illustrations with Kondo impurity model}\label{sec:num}
% \subsection{Physical models and their Hamiltonian\label{2B}

%\subsection{Kondo impurity model}\label{thsec:model}

The Kondo model considers the interactions between a localized spin--$\frac{1}{2}$ impurity and conduction electrons. The Hamiltonian reads \cite{Hew93}
\begin{align}
  H_{\tK} = h_{\B} + H_{\rm int},
\end{align}
where the interaction takes the generic exchange interaction form,
\begin{align}\label{exc}
  H_{\rm int} &= \frac{J}{2}\Big[ \hat S^{\rm imp}_z(\hat\Phi^+_{\uparrow}\hat\Phi^-_{\uparrow}-\hat\Phi^+_{\downarrow}\hat\Phi^-_{\downarrow}) \nl 
  &\quad +\hat S^{\rm imp}_{-}\hat\Phi^+_{\uparrow}\hat\Phi^-_{\downarrow}+\hat S^{\rm imp}_+ \hat\Phi^+_{\downarrow}\hat\Phi^-_{\uparrow}\Big],
\end{align}
with $\hat{\bm S}^{\rm imp} \equiv \frac{1}{2}\sum_{ss'}\hat a^+_{s}\hat{\bm\sigma}_{ss'}\hat a_{s'}^-$ being the impurity spin operators expressed in terms of system creation and annihilation operators, $J$ being the coupling constant. 
{\color{black}
Here, $\hat S_{\pm} \equiv \hat S_x\pm i\hat S_y$ and $\hat{\bm\sigma} \equiv (\hat\sigma_x,\hat\sigma_y,\hat\sigma_z)$ are the Pauli matrices. 
The $(\hat\sigma_{i})_{ss'}$ is the element in $s$-row and $s'$-column of the Pauli matrix $\hat \sigma_{i}$ with $i=x,y,z$,\cite{Mah00} for example,
\begin{align}
  (\hat\sigma_z)_{ss'} = 
  \begin{pmatrix}
    1 & 0 \\
    0 & -1
  \end{pmatrix}.
\end{align}
} 

To proceed, we recast \Eq{exc} as 
\begin{align}\label{exc1}
  H_{\rm int} = \frac{1}{2}\sum_{s}\{\hat\Phi^+_s,\hat\Phi^-_{s}\}\hat q^{-+}_{ss} + \frac{1}{2}\sum_{\sigma s,\sigma's'}\hat q^{\bar\sigma\bar\sigma'}_{ss'}\hat\Phi^\sigma_s\hat\Phi^{\sigma'}_{s'},
\end{align}
by denoting 
\begin{align}
  \hat q^{-+}_{ss'} = \hat Q_{ss'},\ \hat q^{+-}_{ss'} = -\hat Q_{s's},\ \hat q^{++}_{ss'} = \hat q^{--}_{ss'} = 0
\end{align}
and 
\begin{align}
  \hat{\bf Q} \equiv \frac{J}{2}
  \begin{pmatrix}
    S_z^{\rm imp} & S_-^{\rm imp}\\ 
    S_+^{\rm imp} & -S_z^{\rm imp}
  \end{pmatrix} \equiv (\hat Q_{ss'}).
\end{align}
Since $\{\hat\Phi^+_s,\hat\Phi^-_{s}\} = \sum_{k}|c_{ks}|^2$ is a c--number, the first term in \Eq{exc1} is just a system subspace operator. In this sense, the Kondo model can be written as the quadratic system--bath composite Hamiltonian, namely, 
\begin{align}
  H_{\tK} &= \frac{1}{2}\sum_{s}\{\hat\Phi^+_s,\hat\Phi^-_{s}\}\hat q^{-+}_{ss} + h_{\B} + \frac{1}{2}\sum_{\sigma s,\sigma's'}\hat q^{\bar\sigma\bar\sigma'}_{ss'}\hat\Phi^\sigma_s\hat\Phi^{\sigma'}_{s'} \nl 
  &\equiv H_{\tS} + h_{\B} + H_{\SB}.
\end{align}

%\subsection{Spectral function}

For the Kondo model, the spin spectral function is defined via \cite{Hew93}
\begin{align}\label{eq33}
  A_s(\omega) \equiv \frac{1}{2\pi}\int_{-\infty}^\infty\!\!\ud t\,e^{i\omega t}\la \{\delta\hat O_{s}(t),\delta\hat O^\dagger_s(0) \} \ra,
\end{align}
with $\delta\hat O_s \equiv \hat O_s - \la\hat O_s\ra$, 
\begin{align}
  \hat O_s \equiv -\sum_{s'}\hat q^{+-}_{s's}\hat\Phi^{-}_{s'} = \sum_{s'}\hat Q_{ss'}\hat\Phi_{s'}^-,
\end{align}
and 
\begin{align}
  \hat O^\dagger_s = \sum_{s'}\hat q^{-+}_{s's}\hat\Phi^{+}_{s'} = \sum_{s'}\hat Q_{s's}\hat\Phi_{s'}^+.
\end{align}
Here, the average is evaluated with respect to the steady state of the total system. 
{\color{black}
See Ref.\,\onlinecite{Yan16110306} for the algorithm  evaluating $\la \{\delta\hat O_{s}(t),\delta\hat O^\dagger_s(0) \} \ra$ in \Eq{eq33}. 
}
The impurity spectral function is defined as
\begin{align}
  A(\omega) \equiv \sum_{s=\uparrow,\downarrow}A_s(\omega).
\end{align}

%\subsection{Numerical illustration}

%
Using the ext-DEOM, we calculate the impurity spectral function at different temperatures.
In the numerical illustration, we model the bath with the Lorentz type spectral function, namely,
{\color{black}
\begin{align}
  \Gamma_{s}^-(\omega) =\pi N(0) \frac{1}{1+(\omega/W)^2} = \Gamma_s^+(\omega)
\end{align}
with $N(\w)$ being the density of state per spin and $W$ the band width. In the numerical simulations, we set $N(0)=2/(\pi W)$.
}
As expected, our results show that at low temperatures, a sharp peak emerges in the Kondo spectrum at the Fermi energy, near $\w=0$, with a width that decreases as the temperature is lowered. 
This peak corresponds to the Kondo resonance, which is a signature of the effective screening of the impurity spin by the conduction electrons.
Overall, our numerical simulations of the Kondo spectral function confirm the existence of the Kondo resonance, exhibiting its dependence on the temperature.
{\color{black}
In these simulations, the number of exponential terms, $K$ in \Eqs{Fc} and (\ref{Fcc}), is 2, 3, 5, 6 and 7 for  $\beta W =$ 1, 4, 100, 400 and 1000, respectively. 
The exponential decomposition is done via the time--domain Prony fitting decomposition
scheme. \cite{Che22221102} 
We set the truncation tier to be $n = 6$, which is tested to ensure the convergence of the DEOM calculations.
}
 These results illustrate the power of the ext-DEOM method for studying strongly correlated electron systems.
{\color{black}
They can be compared with that from other methods such as numerical renormalization group.\cite{Ljubljana}
}

\begin{figure}[]
  \centering
  \includegraphics[width = \columnwidth]{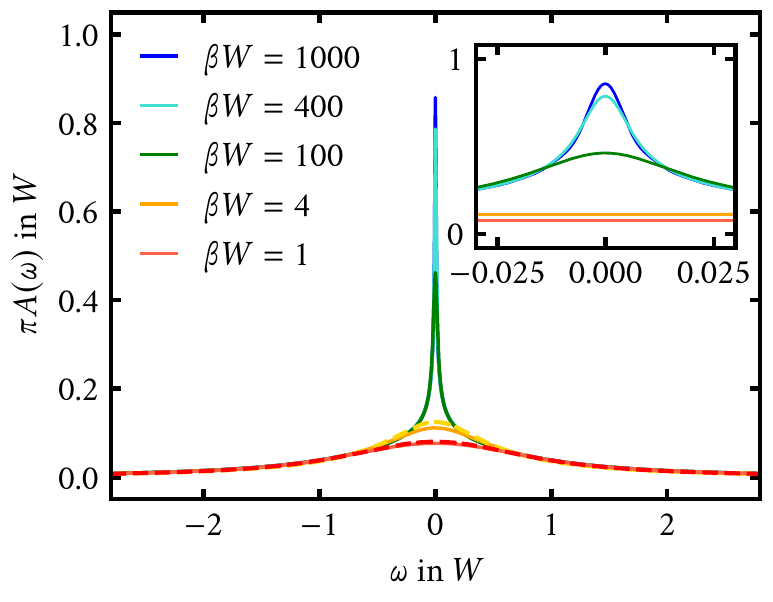}
  \caption{The \textrm{ext}-DEOM simulation results of the impurity spectral function $A(\omega)$ at different temperatures: $\beta W = $ 1, 4, 100, 400 and 1000 with coupling constant $J= 0.3 W$. The perturbative results are plotted in dash lines for $\beta W = 1$ and $4$. Inset is the zooming in on the peaks.
{\color{black}The Friedel sum is $\pi A(\w=0, T=0)=1$ in unit of $W$}.  
  }\label{fig1}
\end{figure}

As shown in \Fig{fig1}, when higher than the Kondo temperature, given by $\beta_{\tK}W \sim 200$, the perturbative results (dash lines) match well with exact ones (solid lines). The former are computed by truncating \Eq{eDEOM} up to tier $n=2$. 
When much lower than $T_{\tK}$, the Kondo temperature, the Kondo peak becomes prominent; see the green, light blue and dark blue  lines in \Fig{fig1}. These lines can not be reproduced quantitatively via perturbative methods. Perturbation  gives rise to much larger spurious peaks. For example, in the case of $\beta W = 100$, it gives $\pi A(0) \sim 2.2W$ (not shown in the figure), which manifestly violates 
{\color{black}
the Friedel sum, $\pi A(\w=0, T=0)=[\pi N(0)]^{-1}=1$ in unit of $W$. \cite{Lan66516, Cos001504}}

\section{Concluding remarks}
\label{sum}

Obtaining and understanding dynamics for quantum impurity system are of great significance in various fields. The DEOM formalism is proposed and used as a standard theoretical framework to describe the dynamics of impurities embedded in environments. In this work, an extended DEOM is presented to deal with quadratic couplings for electronic open quantum systems. The full DEOM formalism offers a powerful tool for studying the noval behaviors in electronic impurity systems and is particularly useful in situations where nonequilibrium and strongly correlated effects are significant. 

Numerical simulations are carried out to investigate the temperature dependence of the Kondo resonance in quantum dots represented by the Kondo model, demonstrating the usefulness of the proposed extension. 
It is anticipated that fermionic ext-DEOM dissipaton theories would become essential towards the characterization of electronic quantum impurities, 
{\color{black}
whose formulations  cover the \Sch picture,  the Heisenberg picture, and further the imaginary--time calculations. \cite{Wan22170901}

Despite these advantages, DEOM faces a huge computational effort in calculating the impurity properties at extremely low temperatures, compared to
other methods that can used to treat the Kondo model,
for example, the numerical renormalization group method. 
 This largely limits the applications of DEOM  in these scenarios.
Many efforts are devoted to alleviate this difficulty; see Refs.\,\onlinecite{Tan20020901} and \onlinecite{Wan22170901} for more information.
}

\section*{Supplemental material}

{\color{black}
\noindent The supplementary material is available at:
\begin{itemize}[leftmargin=*]
  \item The relevant code using in this work can be found in the \textsc{Moscal} 2.0 project (\texttt{fermi-quad} module) at \href{https://git.lug.ustc.edu.cn/czh123/moscal2.0}{https://git.lug.ustc.edu.cn/czh123/moscal2.0}.
\end{itemize}
 }

\begin{acknowledgements}
	Support from the Ministry of Science and Technology of China (Grant No.\ 2021YFA1200103) and the National Natural Science Foundation of China (Grant Nos.\ 22103073 and 22173088) is gratefully acknowledged.
	We  thank  the  USTC  supercomputing  center  for  providing partial computational resources for this project.

\vspace{1em}
\end{acknowledgements}

% \bibliographystyle{./aip}
% \bibliography{./bibrefs}

\begin{thebibliography}{10}

  \bibitem{Kon68437}
  J.~Kondo,
  \newblock Phys. Rev. {\bf 169}, 437 (1968).
  
  \bibitem{Sma83515}
  G.~J. Small,
  \newblock in {\em Spectroscopy and Excitation Dynamics of Condensed Molecular
    Systems}, edited by V.~M. Agranovich and R.~M. Hochstrasser, page 515,
    North-Holland Publishing Company, Amsterdam, 1983.
  
  \bibitem{Zit06045312}
  R.~{\v{Z}}itko and J.~Bon{\v{c}}a,
  \newblock Phys. Rev. B {\bf 74}, 045312 (2006).
  
  \bibitem{Bal06373}
  A.~V. Balatsky, I.~Vekhter, and J.-X. Zhu,
  \newblock Rev. Mod. Phys. {\bf 78}, 373 (2006).
  
  \bibitem{Foo15103112}
  R.~H. Foote, D.~R. Ward, J.~R. Prance, J.~K. Gamble, E.~Nielsen,
    B.~Thorgrimsson, D.~E. Savage, A.~L. Saraiva, M.~Friesen, S.~N. Coppersmith,
    and M.~A. Eriksson,
  \newblock Appl. Phys. Lett. {\bf 107}, 103112 (2015).
  
  \bibitem{Her913720}
  S.~Hershfield, J.~H. Davies, and J.~W. Wilkins,
  \newblock Phys. Rev. Lett. {\bf 67}, 3720 (1991).
  
  \bibitem{Din9715521}
  G.-H. Ding and T.-K. Ng,
  \newblock Phys. Rev. B {\bf 56}, R15521 (1997).
  
  \bibitem{Din97R15521}
  G.-H. Ding and T.-K. Ng,
  \newblock Phys. Rev. B {\bf 56}, R15521 (1997).
  
  \bibitem{Sch9814978}
  A.~Schiller and S.~Hershfield,
  \newblock Phys. Rev. B {\bf 58}, 14978 (1998).
  
  \bibitem{Dic014505}
  N.~L. Dickens and D.~E. Logan,
  \newblock J. Phys.: Condens. Matter {\bf 13}, 4505 (2001).
  
  \bibitem{Luo04256602}
  H.~G. Luo, T.~Xiang, X.~Q. Wang, Z.~B. Su, and L.~Yu,
  \newblock Phys. Rev. Lett. {\bf 92}, 256602 (2004),
  \newblock Reply: {\bf 96}, 019702 (2006).
  
  \bibitem{Hur07126801}
  K.~Le~Hur, P.~Doucet-Beaupr\'e, and W.~Hofstetter,
  \newblock Phys. Rev. Lett. {\bf 99}, 126801 (2007).
  
  \bibitem{Li982893}
  J.~T. Li, W.-D. Schneider, R.~Berndt, and B.~Delley,
  \newblock Phys. Rev. Lett. {\bf 80}, 2893 (1998).
  
  \bibitem{Rui14035119}
  D.~A. Ruiz-Tijerina, E.~Vernek, and S.~E. Ulloa,
  \newblock Phys. Rev. B {\bf 90}, 035119 (2014).
  
  \bibitem{Ujs002557}
  O.~\'Ujs\'aghy, J.~Kroha, L.~Szunyogh, and A.~Zawadowski,
  \newblock Phys. Rev. Lett. {\bf 85}, 2557 (2000).
  
  \bibitem{Ham04115313}
  M.~Hamasaki,
  \newblock Phys. Rev. B {\bf 69}, 115313 (2004).
  
  \bibitem{Sch09235130}
  S.~Schmitt, T.~Jabben, and N.~Grewe,
  \newblock Phys. Rev. B {\bf 80}, 235130 (2009).
  
  \bibitem{Isi10235120}
  A.~Isidori, D.~Roosen, L.~Bartosch, W.~Hofstetter, and P.~Kopietz,
  \newblock Phys. Rev. B {\bf 81}, 235120 (2010).
  
  \bibitem{Coh11075150}
  G.~Cohen and E.~Rabani,
  \newblock Phys. Rev. B {\bf 84}, 075150 (2011).
  
  \bibitem{Tsu11187201}
  N.~Tsukahara, S.~Shiraki, S.~Itou, N.~Ohta, N.~Takagi, and M.~Kawai,
  \newblock Phys. Rev. Lett {\bf 106}, 187201 (2011).
  
  \bibitem{Ort12125324}
  C.~P. Orth, D.~F. Urban, and A.~Komnik,
  \newblock Phys. Rev. B {\bf 86}, 125324 (2012).
  
  \bibitem{Zha17075410}
  Z.~Q. Zhang, S.~Li, J.~T. L\"u, and J.~Gao,
  \newblock Phys. Rev. B {\bf 96}, 075410 (2017).
  
  \bibitem{Gul11349}
  E.~Gull, A.~J. Millis, A.~I. Lichtenstein, A.~N. Rubtsov, M.~Troyer, and
    P.~Werner,
  \newblock Rev. Mod. Phys. {\bf 83}, 349 (2011).
  
  \bibitem{Har15085430}
  R.~H\"artle, G.~Cohen, D.~R. Reichman, and A.~J. Millis,
  \newblock Phys. Rev. B {\bf 92}, 085430 (2015).
  
  \bibitem{Wil75773}
  K.~G. Wilson,
  \newblock Rev. Mod. Phys. {\bf 47}, 773 (1975).
  
  \bibitem{Yos909403}
  M.~Yoshida, M.~A. Whitaker, and L.~N. Oliveira,
  \newblock Phys. Rev. B {\bf 41}, 9403 (1990).
  
  \bibitem{Cos973003}
  T.~A. Costi,
  \newblock Phys. Rev. B {\bf 55}, 3003 (1997).
  
  \bibitem{Bul988365}
  R.~Bulla, A.~C. Hewson, and T.~Pruschke,
  \newblock J. Phys.: Cond. Matt. {\bf 10}, 8365 (1998).
  
  \bibitem{Pet06245114}
  R.~Peters, T.~Pruschke, and F.~B. Anders,
  \newblock Phys. Rev. B {\bf 74}, 245114 (2006).
  
  \bibitem{And08195216}
  F.~B. Anders,
  \newblock J. Phys.: Condens. Matter {\bf 20}, 195216 (2008).
  
  \bibitem{And05196801}
  F.~B. Anders and A.~Schiller,
  \newblock Phys. Rev. Lett. {\bf 95}, 196801 (2005).
  
  \bibitem{Fri06144410}
  L.~Fritz, S.~Florens, and M.~Vojta,
  \newblock Phys. Rev. B {\bf 74}, 144410 (2006).
  
  \bibitem{Nis04613}
  S.~Nishimoto and E.~Jeckelmann,
  \newblock J. Phys.: Condens. Matter {\bf 16}, 613 (2004).
  
  \bibitem{Mer12075153}
  L.~Merker, A.~Weichselbaum, and T.~A. Costi,
  \newblock Phys. Rev. B {\bf 86}, 075153 (2012).
  
  \bibitem{Pri10050404}
  J.~Prior, A.~W. Chin, S.~F. Huelga, and M.~B. Plenio,
  \newblock Phys. Rev. Lett. {\bf 105}, 050404 (2010).
  
  \bibitem{Nus20155134}
  A.~N\"u\ss{}eler, I.~Dhand, S.~F. Huelga, and M.~B. Plenio,
  \newblock Phys. Rev. B {\bf 101}, 155134 (2020).
  
  \bibitem{Jor19240602}
  M.~R. J\o{}rgensen and F.~A. Pollock,
  \newblock Phys. Rev. Lett. {\bf 123}, 240602 (2019).
  
  \bibitem{Ric22167403}
  M.~Richter and S.~Hughes,
  \newblock Phys. Rev. Lett. {\bf 128}, 167403 (2022).
  
  \bibitem{Cyg22662}
  M.~Cygorek, M.~Cosacchi, A.~Vagov, V.~M. Axt, B.~W. Lovett, J.~Keeling, and
    E.~M. Gauger,
  \newblock Nat. Phys. {\bf 18}, 662 (2022).
  
  \bibitem{Coh15266802}
  G.~Cohen, E.~Gull, D.~R. Reichman, and A.~J. Millis,
  \newblock Phys. Rev. Lett. {\bf 115}, 266802 (2015).
  
  \bibitem{Che17054105}
  H.-T. Chen, G.~Cohen, and D.~R. Reichman,
  \newblock The Journal of chemical physics {\bf 146}, 054105 (2017).
  
  \bibitem{Cai202430}
  Z.~Cai, J.~Lu, and S.~Yang,
  \newblock Communications on Pure and Applied Mathematics {\bf 73}, 2430 (2020).
  
  \bibitem{Ber21155104}
  C.~Bertrand, D.~Bauernfeind, P.~T. Dumitrescu, M.~Ma\ifmmode~\check{c}\else
    \v{c}\fi{}ek, X.~Waintal, and O.~Parcollet,
  \newblock Phys. Rev. B {\bf 103}, 155104 (2021).
  
  \bibitem{Sor19043303}
  M.~E. Sorantin, D.~M. Fugger, A.~Dorda, W.~von~der Linden, and E.~Arrigoni,
  \newblock Phys. Rev. E {\bf 99}, 043303 (2019).
  
  \bibitem{Fey63118}
  R.~P. Feynman and F.~L. \mbox{Vernon, Jr.},
  \newblock Ann. Phys. {\bf 24}, 118 (1963).
  
  \bibitem{Tan20020901}
  Y.~Tanimura,
  \newblock J. Chem. Phys {\bf 153}, 020901 (2020).
  
  \bibitem{Tan89101}
  Y.~Tanimura and R.~Kubo,
  \newblock J. Phys. Soc. Jpn. {\bf 58}, 101 (1989).
  
  \bibitem{Tan906676}
  Y.~Tanimura,
  \newblock Phys. Rev. A {\bf 41}, 6676 (1990).
  
  \bibitem{Yan04216}
  Y.~A. Yan, F.~Yang, Y.~Liu, and J.~S. Shao,
  \newblock Chem. Phys. Lett. {\bf 395}, 216 (2004).
  
  \bibitem{Tan06082001}
  Y.~Tanimura,
  \newblock J. Phys. Soc. Jpn. {\bf 75}, 082001 (2006).
  
  \bibitem{Xu05041103}
  R.~X. Xu, P.~Cui, X.~Q. Li, Y.~Mo, and Y.~J. Yan,
  \newblock J. Chem. Phys. {\bf 122}, 041103 (2005).
  
  \bibitem{Xu07031107}
  R.~X. Xu and Y.~J. Yan,
  \newblock Phys. Rev. E {\bf 75}, 031107 (2007).
  
  \bibitem{Din12224103}
  J.~J. Ding, R.~X. Xu, and Y.~J. Yan,
  \newblock J. Chem. Phys. {\bf 136}, 224103 (2012).
  
  \bibitem{Jin08234703}
  J.~S. Jin, X.~Zheng, and Y.~J. Yan,
  \newblock J. Chem. Phys. {\bf 128}, 234703 (2008).
  
  \bibitem{Li12266403}
  Z.~H. Li, N.~H. Tong, X.~Zheng, D.~Hou, J.~H. Wei, J.~Hu, and Y.~J. Yan,
  \newblock Phys. Rev. Lett. {\bf 109}, 266403 (2012).
  
  \bibitem{Ye16608}
  L.~Z. Ye, X.~L. Wang, D.~Hou, R.~X. Xu, X.~Zheng, and Y.~J. Yan,
  \newblock WIREs Comp. Mol. Sci. {\bf 6}, 608 (2016).
  
  \bibitem{Cos001504}
  T.~A. Costi,
  \newblock Phys. Rev. Lett. {\bf 85}, 1504 (2000).
  
  \bibitem{Yan14054105}
  Y.~J. Yan,
  \newblock J. Chem. Phys. {\bf 140}, 054105 (2014).
  
  \bibitem{Wan22170901}
  Y.~Wang and Y.~J. Yan,
  \newblock J. Chem. Phys. {\bf 157}, 170901 (2022).
  
  \bibitem{Xu17395}
  R.~X. Xu, Y.~Liu, H.~D. Zhang, and Y.~J. Yan,
  \newblock Chin. J. Chem. Phys. {\bf 30}, 395 (2017).
  
  \bibitem{Xu18114103}
  R.~X. Xu, Y.~Liu, H.~D. Zhang, and Y.~J. Yan,
  \newblock J. Chem. Phys. {\bf 148}, 114103 (2018).
  
  \bibitem{Che23074102}
  Z.-H. Chen, Y.~Wang, R.-X. Xu, and Y.~Yan,
  \newblock J. Chem. Phys. {\bf 158}, 074102 (2023).
  
  \bibitem{Yan16110306}
  Y.~J. Yan, J.~S. Jin, R.~X. Xu, and X.~Zheng,
  \newblock Frontiers Phys. {\bf 11}, 110306 (2016).
  
  \bibitem{Bog09}
  N.~N. Bogoliubov and N.~N. Bogoliubov~Jr,
  \newblock {\em Introduction to Quantum Statistical Mechanics (Second Edition)},
  \newblock World Scientific Publishing Company, 2009.
  
  \bibitem{Hew93}
  A.~C. Hewson,
  \newblock {\em The Kondo Problem to Heavy Fermions},
  \newblock Cambridge University Press, Cambridge, 1993.
  
  \bibitem{Mah00}
  G.~D. Mahan,
  \newblock {\em Many-Particle Physics},
  \newblock Plenum, New York, 3rd edition, 2000.
  
  \bibitem{Che22221102}
  Z.~H. Chen, Y.~Wang, X.~Zheng, R.~X. Xu, and Y.~J. Yan,
  \newblock J. Chem. Phys. {\bf 156}, 221102 (2022).
  
  \bibitem{Ljubljana}
  {Zitko, Rok.},
  \newblock Nrg ljubljana.
  
  \bibitem{Lan66516}
  D.~C. Langreth,
  \newblock Phys. Rev. {\bf 150}, 516 (1966).
  
  \end{thebibliography}

\end{document}